\documentclass[aps,prc,twocolumn,showpacs]{revtex4-1}

\usepackage{graphicx} 
\usepackage{dcolumn} 
\usepackage{bm}         

\newcommand{\be}{\begin{equation}}
\newcommand{\ee}{\end{equation}}
\newcommand{\ba}{\begin{eqnarray}}
\newcommand{\ea}{\end{eqnarray}}
\newcommand{\bd}{\begin{displaymath}}
\newcommand{\ed}{\end{displaymath}}

\begin{document}

\title{QGP flow fluctuations and the caracteristics of higher moments}

\author{
D. J. Wang$^{1,2}$,
L. P. Csernai$^{1}$,
D. Strottman$^1$,
Cs. Anderlik$^3$,
Y. Cheng$^2$,\\
D. M. Zhou$^2$,
Y. L. Yan$^4$,
X. Cai$^2$
and
B. H. Sa$^4$,
}

\affiliation{
${1}$ Department of Physics and Technology, University of Bergen, 5007 Bergen, Norway\\
${2}$ Institute of Particle Physics, Huazhong Normal University, 430079 Wuhan, China\\
${3}$ Uni Computing, Thorm{\o}hlensgate 55, N-5008 Bergen, Norway\\
${4}$ China Institute of Atomic Energy, P. O. Box 275 (10), 102413 Beijing, China
 }

\date{\today}

\begin{abstract}
The dynamical development of expanding Quark-gluon Plasma (QGP) flow is
studied in a 3+1D fluid dynamical model with a globally symmetric, initial
condition.  We minimize fluctuations arising from complex dynamical
processes at finite impact parameters and from fluctuating random
initial conditions to have a conservative fluid dynamical background
estimate for the statistical distributions of the thermodynamical
parameters. We also avoid a phase transition in the equation of state,
and we let the matter supercool during the expansion.
  Then central Pb+Pb collisions at $\sqrt{s_{NN}} = 2.76$ TeV are
studied in an almost perfect fluid dynamical model, with azimuthally
symmetric initial state generated in a dynamical flux-tube model.
The general development of thermodynamical extensives are also shown
for lower energies.
   We observe considerable deviations from a thermal equilibrium
source, changing skewness and kurtosis by time depending on beam energy
as a consequence of the fluid dynamical expansion arising from a
least fluctuating initial state.
\end{abstract}




\maketitle
\section{Collective Flow of Global Symmetry}
\label{intro}
In heavy ion collisions
collective flow has been measured and azimuthal asymmetry
was determined from $v_1$ to $v_8$.
At the highest energies in central collisions fluctuations
dominate arising from fluctuating initial configurations, and the most
dominant flow harmonic is $v_3$. These are collective flow fluctuations
and have no direct connection to the fluctuations arising from a
phase transition in the Equation of State (EoS) \cite{Stephano1,Stephano2,Stephano3,PRCZhou}.

The fluctuations arising from the pure fluid
dynamics without hadronization are studied in this work.
We present a set of calculations,
what kind of effects a possibly most conservative relativistic fluid
dynamical model exhibits in higher moments of statistical parameters
for extensive densities, as skewness and kurtosis, without including
any effect associated with a phases transition, in the EoS, in the
transport properties, or in special thermodynamical phase space
trajectories \cite{Bass1,Bass2,Bass3},
or in special freeze out mechanisms. This study complements
a large number of studies with the opposite goal, aiming to analyze
the consequences of the above mentioned effects. Contrary to the effort
to eliminate all these effects, we still obtain energy dependent changes
in the skewness and kurtosis.

We use an EoS without a phase transition and include the possibility
of supercooling: we choose the MIT Bag model with parameters fixed to
the initial values (two flavors and massless
quarks and gluons, and the bag constant is $ B=0.397$ GeV/$fm^3$).
In addition to omitting the freeze out and its effects,
we also avoid to take into account viscosity (except the inavoidable
numerical viscosity), thus also the temperature dependence
of viscosity near the critical point \cite{Niemi}, which may lead to
additional changes of the critical fluctuations in a viscous fluid
dynamical evolution.

In peripheral collisions, on the other hand, the initial asymmetry is
dominated by the almond shape of the participant matter. This results
in a strong elliptic flow, while the directed flow measured in the
ALICE TPC appeared to be weak and dominated by random fluctuations.

Computational Fluid Dynamics (CFD) predictions indicated a new directed flow
structure:
due to the large angular momentum of the initial state in peripheral
collisions the anti-flow peak observed at high SPS and RHIC energies
is rotating forward, and
at sufficiently high beam energy,
 $v_1(y)$ will start to peak at positive rapidities, i.e.
on the same side where the projectile spectator residues arrive after the
collision \cite{CsMSS2011}. This happens because the initial angular momentum
leads to a faster rotating initial system, and this rotation moves the
dominant directed flow peak forward before the expansion from the
pressure would slow down the rotation. The observation of this peak
is not easy because of the beam directed fluctuations of the initial
state.

At high energies the dimensionless shear viscosity over the entropy
density, $\eta / s$, of the QGP is becoming small \cite{Kovtun2005}, and
$\eta/s$ as a function of temperature has a minimum
at the critical temperature \cite{CKM}. So the Reynolds number
may exceed one, and turbulent phenomena may start to occur.
On the other hand, $\eta/s$ at a critical point does not
necessarily have a minimum, since the dynamical universality
class of a possible critical point of QCD is the H-model in
Hohenberg and Halperin's classification \cite{teaney09,Hohenberg},
the shear viscosity may diverge at a possible QCD critical point.
This would damp instabilities.

Recently in the same CFD model with the Particle in Cell (PIC) method,
it was observed that in peripheral collisions
with low viscosity (and low numerical viscosity) a Kelvin-Helmholtz (KH)
instability starts to develop \cite{CSA11}, which enhances the rotation
effect and the spatial variance of the flow pattern.
The flow effects depend on the initial state profile and on the
viscosity. Turbulence appears only for small viscosity, which indicates
the critical point of the matter \cite{CKM}, and it is a sensitive
measure of viscosity and its minimum at the critical point.

Our CFD simulations of the LHC heavy ion collisions
suggest that collective
directed $v_1(y)$ flow function can be measured if the Globally
asymmetric flow component and the random flow arising
from the initial state $y_{CM}$-fluctuations can be separated. In
hydrodynamical calculations we see
that the $v_1$ Global flow can change the peak position
to "forward" with increasing beam energy and initial
angular momentum.   This is a result of our tilted initial
state with shear flow \cite{MCs001,MCs002}, in which the angular
momentum from the increasing beam momentum may supersede the
expansion driven by the pressure.

The above described phenomena contribute to an increased spatial spread
of the matter during the collision. These effects are present even if
we do not have a phase transition in our EoS \cite{NPANayak}.
Our present goal is
to determine the lowest possible deviation from an ideal thermal
source in a least fluctuating CFD evolution.  Thus, we eliminate
random initial state fluctuations and all azimuthal asymmetries
in a head-on collisions to obtain a most symmetric
distribution, and study the spatial fluctuation of thermodynamical
quantities in such a system.  This then can be compared to the
effects caused by the phase transition
\cite{Stephano1,Stephano2,Stephano3,Bass1,Bass2,Bass3,PRDGavai,cser,laszlo1204}.

Most random fluctuations lead to close to Gaussian distribution,
nevertheless the dominant
fluid dynamical expansion, even if all special sources are
eliminated or minimized may lead
to more complex non-gaussian fluctuations and higher statistical moments.

For a realistic reaction model we have to describe the final
stage of the reaction also.
We have a Multi Module Model approach to describe high energy
heavy ion collisions in the RHIC and LHC energy range.
Then from the locally equilibrated QGP we have to form hadrons.
We do not assume that the hadronization happens in chemical equilibrium
as this would take too long time \cite{CK921,CK922} and would not allow for
baryons of high strangeness. Thus we use the simplest bag model approach,
which in the pure QGP domain, yields similar results to more detailed
parametrizations fitted to lattice predictions \cite{Huovinen}.
\begin{figure}[ht] 
\begin{center}
\resizebox{0.9\columnwidth}{!}
{\includegraphics{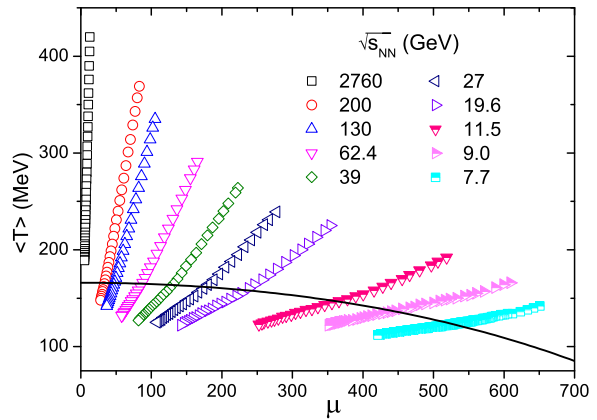}}
\caption{(color online) The trajectory of fluid dynamical
development of QGP fluid at different
beam energies as indicated in the figure. Open squares belong to
central Pb+Pb collisions,
the others are central Au+Au collisions, these are
calculated at a cell size resolution of $dx=dy=dz=0.575$ fm, and time step
$\Delta t = 0.04233$ fm/c. The hadronic freeze out curve \cite{Cleymans}
is indicated by a full black line. The CFD evolution is calculated well
beyond this curve. This is possible as the CFD model can describe supercooled
QGP fluid also. The viscosity is minimal and only the numerical viscosity
is considered in the calculations. These can be performed down to
FAIR and NICA energies, although the use of supercooled QGP EoS has
constrained validity at these low energies. }
\label{fig1}
\end{center}
\end{figure}

In order to be able to hadronize rapidly we have to assume
a fast, non-equilibrium hadronization and freeze out with either
a Cooper-Frye based method with a local sudden change or a sudden
transition to a parton and hadron cascade model, e.g. \cite{PACIAESa},
which can describe rapid hadronization without the assumption of
chemical equilibration.
This final stage should additionally increase the deviations
from the local statistical equilibrium. Thus, the final observed
(or calculated) particle distribution, which contains already the
influence of the rapid hadronization and freeze out, should be compared
to the basic fluctuations arising from a (least fluctuating) CFD
distribution estimate.

It is also important to mention that different thermodynamical
parameters (especially intensives and extensives) do not have to
show the same critical fluctuation properties, so we have to
study the fluctuations of several parameters. Furthermore the
statistical physics estimates assume a single thermal source
at or near the critical point, while we estimate here also the
effects of spatial fluctuations, which arise from a dynamically
expanding fluid flow even in the least fluctuating configuration.
\begin{figure}[ht] 
\begin{center}
\resizebox{0.9\columnwidth}{!}
{\includegraphics{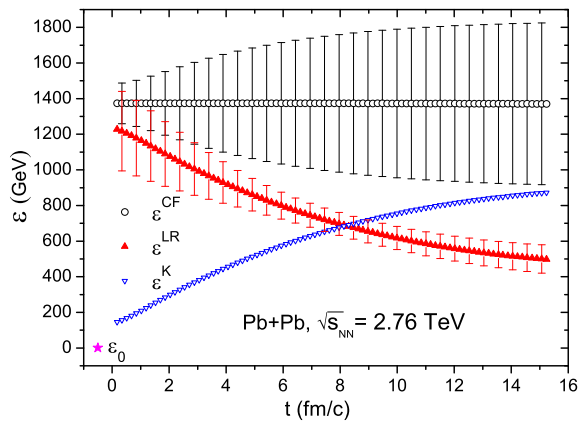}}
\caption{ (color online) The average specific energy density in the
calculational frame ($\varepsilon^{CF}$), the average specific internal
energy density in the local rest frame ($\varepsilon^{LR}$) and the average
specific kinetic energy
($\varepsilon^K = \varepsilon^{CF} - \varepsilon^{LR} $) are calculated by
the PIC hydro model in central collisions at $\sqrt{s_{NN}}=2.76$ TeV. The
star indicates $\varepsilon_0=0.938$ GeV, the initial specific internal
energy before collision. The error bars indicate the variance, $\sigma$.}
\label{fig2}
\end{center}
\end{figure}
\begin{figure}[ht] 
\begin{center}
\resizebox{0.9\columnwidth}{!}
{\includegraphics{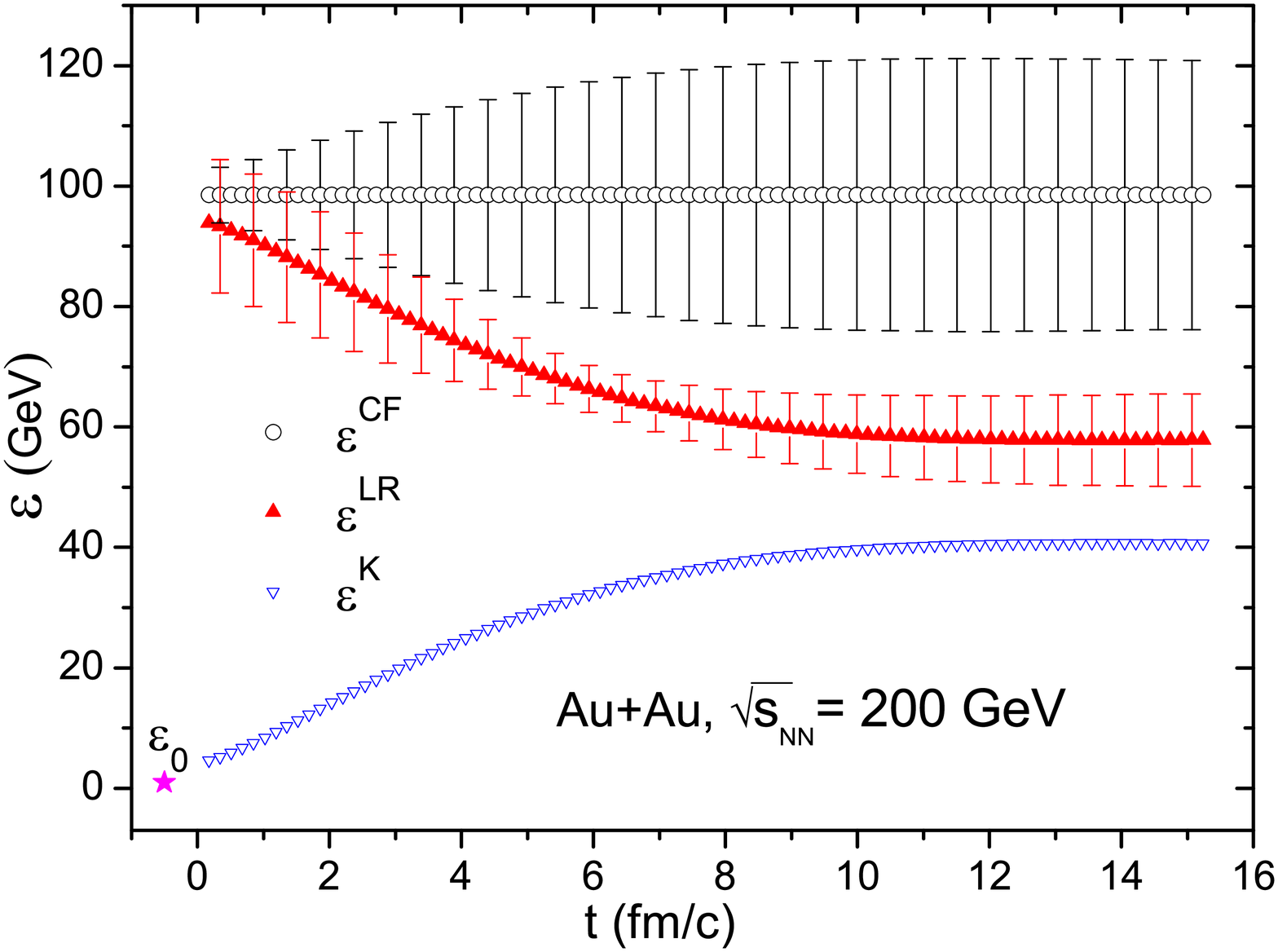}}
\caption{ (color online) The average specific energy density in the
calculation frame ($\varepsilon^{CF}$), the average specific internal
energy density in the local rest frame ($\varepsilon^{LR}$) and the average
specific kinetic energy ($\varepsilon^K$) are calculated by the PIC hydro
model in central collisions at $\sqrt{s_{NN}}=200$ GeV. $\varepsilon_0=0.938$ GeV
is the initial internal energy before collision. The kinetic energy is the
difference between $\varepsilon^{CF}$ and $\varepsilon^{LR}$.}
\label{fig3}
\end{center}
\end{figure}
\begin{figure}[ht]   
\begin{center}
\resizebox{0.9\columnwidth}{!}
{\includegraphics{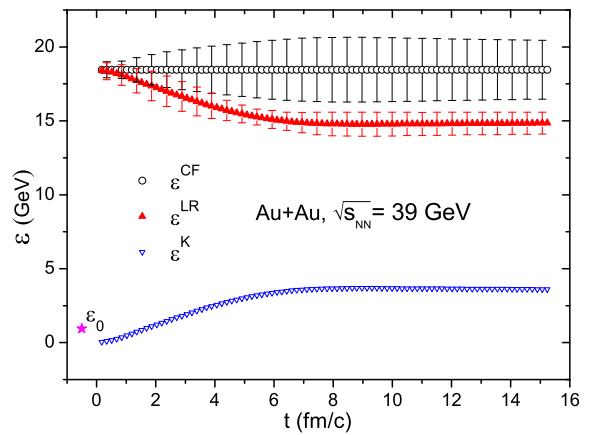}}
\caption{(color online) The average specific energy density in the
calculation frame ($\varepsilon^{CF}$), the average specific internal energy
density in the local rest frame ($\varepsilon^{LR}$) and the average specific
kinetic energy ($\varepsilon^K$) are calculated by the PIC hydro model in
central collisions at $\sqrt{s_{NN}}=39$ GeV. $\varepsilon_0=0.938$ GeV is the
initial internal energy before collision. The kinetic energy is the difference
between $\varepsilon^{CF}$ and $\varepsilon^{LR}$. Observe that at 7 fm/c
the specific internal energy and kinetic energy change stalls indicating that
the majority of the cells reached zero pressure, the FO boundary. This
coincides with the crossing point of the FO curve in Fig. \ref{fig1},
indicating that beyond that point the assumed development in the supercooled
QGP is overstretching the applicability of the CFD calculation.}
\label{fig4}
\end{center}
\end{figure}

\section{Flow development}
\label{sec:1}
The dynamically developing flow pattern leads to a spatial distribution
of all thermodynamical quantities, while the system expands rapidly.
We assume that the most probable scenario is a pre-equilibrium development
described by linear flux-tube expansion independently at each point
of the transverse, [$x,y$], plane until local equilibrium is
reached at a space-time hypersurface. By this time at high energies we
reach a (nearly) equilibrated Quark-gluon Plasma (QGP) state, which
then expands and supercools. This intermediate stage is described with
a CFD model using the PIC method.
Finally the supercooled QGP can hadronize
rapidly and almost simultaneously it freezes out. This final
stage of the reaction can be described by a non-equilibrium model.

The CFD model is using $N$ fluid cells, $i = 1, 2, ... , N$, where
with time and expansion the number of fluid cells is increasing.
Thus, each cell carries less and less baryon charge with time, and
a different amount. Therefore to calculate the volume average and
the distribution over the volume we weight the fluid cells by the
amount of baryon charge they carry.

Thus the weighted average of a quantity $x$ is defined as
\ba
<x> \equiv \sum_i x_i \cdot w_i  \ , \ \ {\rm  where} \ \ \
w_i=\frac{{\sf n}_i^{CF}\cdot V_i}{N_{tot}} \ ,
\ea
and
$V_i$ is the volume of $i^{th}$ fluid cell,
${\sf n}_i^{CF}$ is the baryon density in the Calculational Frame
(CF), which is the c.m. frame in the present calculations,
and
${\sf n}_i^{CF} = n_i^{LR} \gamma_i$ (where $n_i^{LR}$ is the baryon density
in Local Rest Frame), so that
$N_{tot} = \sum_i n_i^{LR} \gamma_i V_i$,
therefore $\sum_i w_i = 1$.

The CFD stage of this development is shown in Fig.~\ref{fig1},
where the time development of the average thermodynamical quantities
is shown in the temperature, $T$ and baryon chemical potential, $\mu$,
plane and the full black line is the hadronic freeze out curve.
In the present work we analyze the CFD stage, and analyze statistically
the space-time development of both the
intensive and specific extensive thermodynamical variables.
We study central collisions only, to avoid the effects from azimuthal
flow asymmetries and from particle emission from projectile and target
residues (spectator evaporation) \cite{Amelin}.

For a variable $x$ the averages and various order moments of cell-by-cell
distributions can be written as
\be
<x^{n}>=\int{x^n P(x) dx} = \sum_i x_i^n \;  w_i  \ ,
\ee

\ba
M^{(n)} &=& <(x-<x>)^n>=\int{(x-<x>)^n P(x) dx}
\nonumber\\
 &=& \sum_i \; (x-<x>)^n \; w_i \ ,
\ea
where $P(x)$ is the spatial distribution weighted
by the baryon charge density in the CF.
The spatial variance, the skewness and the kurtosis can be obtained
from these moments:
\begin{eqnarray}
\sigma^2 = <(x-<x>)^2 = M^{(2)} \ ,
\label{variance}
\end{eqnarray}
\begin{eqnarray}
S = \frac{<(x-<x>)^3>}{\sigma^3} = \frac{M^{(3)}}{(M^{(2)})^{3/2}} \ ,
\label{skewness}
\end{eqnarray}
\begin{eqnarray}
\kappa = <\frac{(x-<x>)^4}{\sigma^4}>-3 =\frac{ M^{(4)}}{(M^{(2)})^2}-3 \ .
\label{kurtosis}
\end{eqnarray}

By using this average, first we can calculate specific extensives,
which are governed by strict conservation
laws. The total baryon charge, energy and momentum conservations
are governed by the continuity
equation and by the relativistic Euler equation.
\ba
N^\mu,_\mu &=& 0 , \\
T^{\mu\nu},_\nu &=& 0 ,
\ea
and as a consequence, the total momentum in the center of mass (c.m.)
frame should remain zero during the development,
while the average specific energy per net nucleon number
\be
<\varepsilon^{CF}> \equiv \frac{T^{00}}{N^0} = const.
\ee
should remain constant in CF.

In most experiments the observable is the total charged particle multiplicity.
This number is proportional to the energy density and not the net baryon charge.
That is the reason we study the energy density. The energy is characterized by
the "specific energy density", i.e., by the energy per unit net baryon charge.
As you can see, the different parts of the energy development undergo significant
changes with time, which may contribute to the statistical properties of the
produced particles.

The average specific energy density in CF
can be expressed as:
\begin{eqnarray}
<\varepsilon^{CF}>=
\sum_i\frac{e_i^{CF}}{{\sf n}_i^{CF}}
\frac{{\sf n}_i^{CF} V_i}{N_{tot}}=
\sum_i\frac{e_i^{CF}V_i}{N_{tot}}  .
\end{eqnarray}
Similarly we can obtain the average specific energy density
in the local rest frame (LR):
\begin{eqnarray}
<\varepsilon^{LR}>=
\sum_i\frac{e_i^{LR}}{n_i^{LR}}
\frac{{\sf n}_i^{CF} V_i}{N_{tot}}=
\sum_i\frac{e_i^{LR} \gamma_i V_i}{N_{tot}} .
\end{eqnarray}

\begin{figure}[ht]   
\begin{center}
\resizebox{0.9\columnwidth}{!}
{\includegraphics{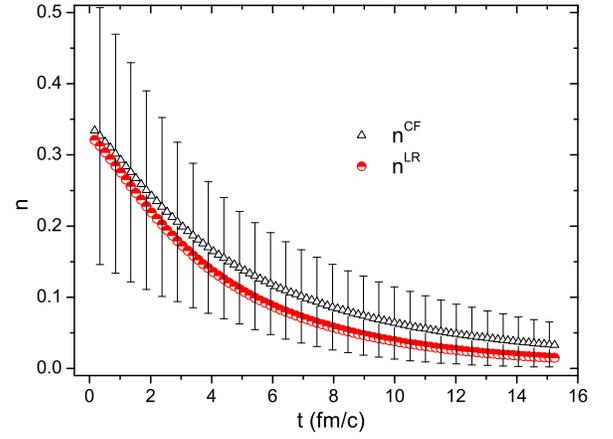}}
\caption{ (color online) The average baryon charge density in the
calculation frame (${\sf n}^{CF}$) and in the local rest frame
($n^{LR}$) calculated by the
PIC hydro model in central Pb+Pb collisions at $\sqrt{s_{NN}}=2.76$ TeV.
The error bars indicate the variance, $\sigma$, for ${\sf n}^{CF}$.
The variance is about the same for the invariant specific density,
$n^{LR}$.
 }
\label{fig5}
\end{center}
\end{figure}

These quantities are shown in Figs.~\ref{fig2},\ref{fig3},\ref{fig4}.
Most importantly the average specific energy remains
constant during the time development, and its value is
the initial beam energy,
$
<\varepsilon^{CF}> = \frac{1}{2} \sqrt{s_{NN}}=1.38
$
TeV.
Due to the
CFD expansion, the average specific internal energy,
$\varepsilon^{LR} = e / n$, decreases as the system expands,
while the expansion leads to increased average specific kinetic energy,
$
\varepsilon^{K} \equiv \varepsilon^{CF} - \varepsilon^{LR} .
$
Of course these quantities vary in the space-time during the CFD
evolution. Their spatial variances, $\sigma_x$, are shown in
eq.~(\ref{variance}), is also indicated
by the error bars in  Figs.~\ref{fig2},\ref{fig3},\ref{fig4}.
Although the average of $\varepsilon^{CF}$
remains constant its variance is increasing due to the expansion,
which generates increasing number of low density cells. The lower
energies give similar results.  The three figures demonstrate how much
part of the energy is converted into kinetic energy. At the highest energy
by the time of 8 fm/c, half of the total available is converted into flow
while at lowest shown beam energy, at $\sqrt{s_{NN}}=39$ GeV, it is only 20\%.

Similarly, the net baryon number, $N_{tot} = \sum_i n_i^{LR} \gamma_i V_i$,
is exactly conserved in the calculations, as the marker particles carry
fixed baryon charge and these are conserved until they are in
the calculation grid. At the same time
the average baryon charge density is decreasing. We can characterize this
by the invariant scalar, LR, baryon density as well as by the CF baryon
density.
Their averages can be expressed as:
\begin{eqnarray}
<{\sf n}^{CF}>=
\sum_i {\sf n}_i^{CF} \;
\frac{{\sf n}_i^{CF} V_i}{N_{tot}}=
\sum_i  ({\sf n}_i^{CF})^2 \;  \frac{ V_i}{N_{tot}}  .
\end{eqnarray}
Similarly we can obtain the average of the invariant scalar baryon density
\begin{eqnarray}
<n^{LR}>=
\sum_i n_i^{LR} \;
\frac{{\sf n}_i^{CF} V_i}{N_{tot}}=
\sum_i  ( n_i^{LR})^2 \; \frac{\gamma_i V_i}{N_{tot}} .
\end{eqnarray}
\begin{figure}[ht]   
\begin{center}
\resizebox{0.9\columnwidth}{!}
{\includegraphics{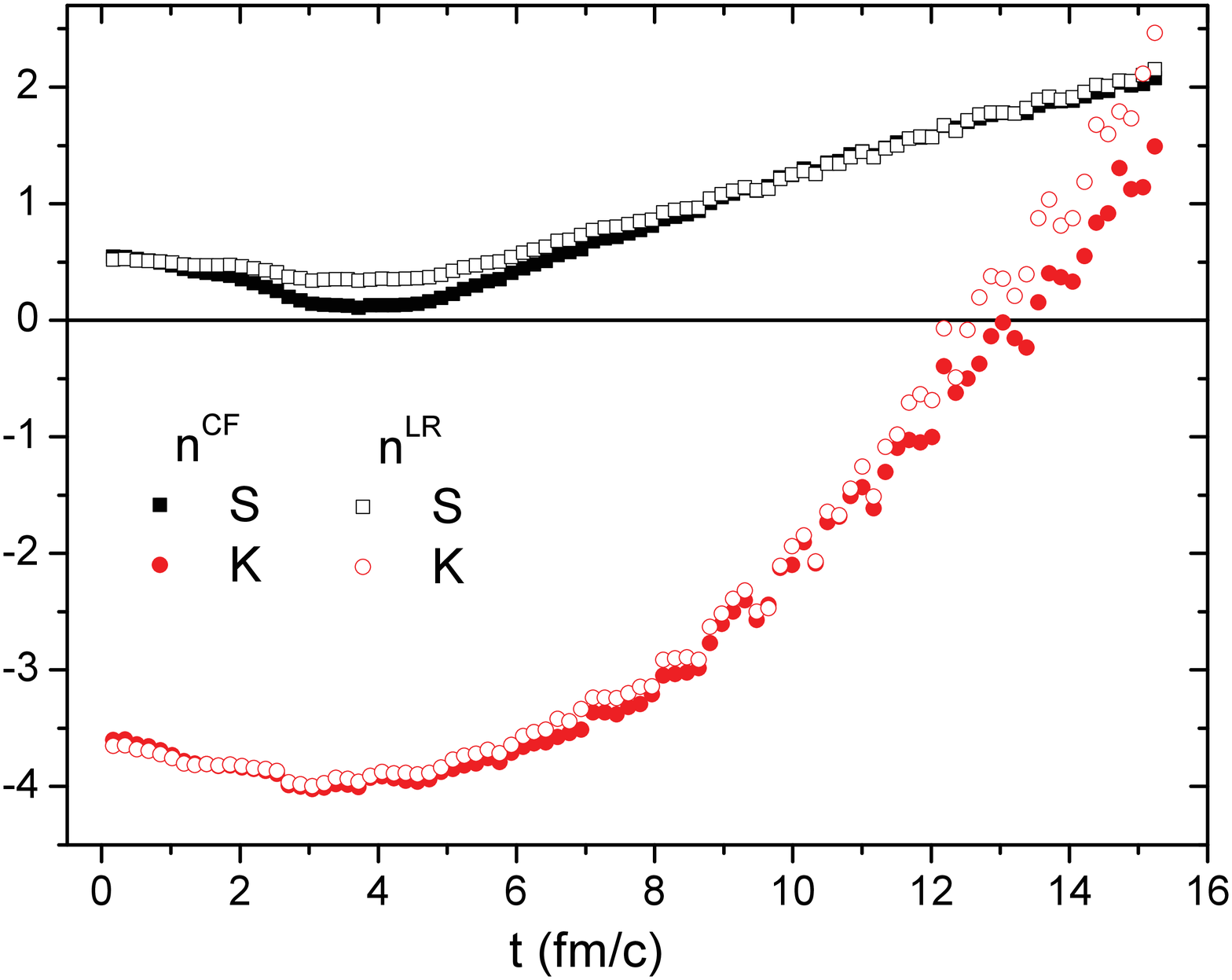}}
\caption{ (color online)
The time development of the skewness and kurtosis
of the distribution of the net baryon density,
${\sf n}^{CF}$, $n^{LR}$, for central
Pb+Pb collision at $\sqrt{s_{NN}}=2.76$ TeV.
The skewness of the baryon charge is positive for both densities
indicating an increasingly longer high density tail of the distribution.
The kurtosis is initially negative, and turns positive only at
very late stages. This is arising from the initial condition where the
baryon charge is uniformly distributed in each longitudinal "streak".
Around 4 fm/c ${\sf n}^{CF}$ is visibly smaller than  $n^{LR}$ indicating
that the apparent density in the CF is more uniform than the invariant
scalar density.
}
\label{fig6}
\end{center}
\end{figure}

The time dependencies of the two densities are shown in Fig.~\ref{fig5}.
We observe that both densities decrease with time, and
the density calculated in CF is larger than the density
calculated in the LR. Their variance is decreasing with time.

\subsection{Late stages of expansion}
\label{sec:2}
As mentioned above the EoS of our perfect fluid dynamical model
is ideal QGP, in the form of the MIT Bag model. The parameter of the
Bag constant is fixed until the local pressure is positive. In order
to be able to calculate the continued expansion in the supercooled QGP
we set and fix the pressure to zero and decrease the Bag constant so that the
expansion in the supercooled state remains adiabatic \cite{Horvat}.
\begin{figure}[ht]   
\begin{center}
\resizebox{0.9\columnwidth}{!}
{\includegraphics{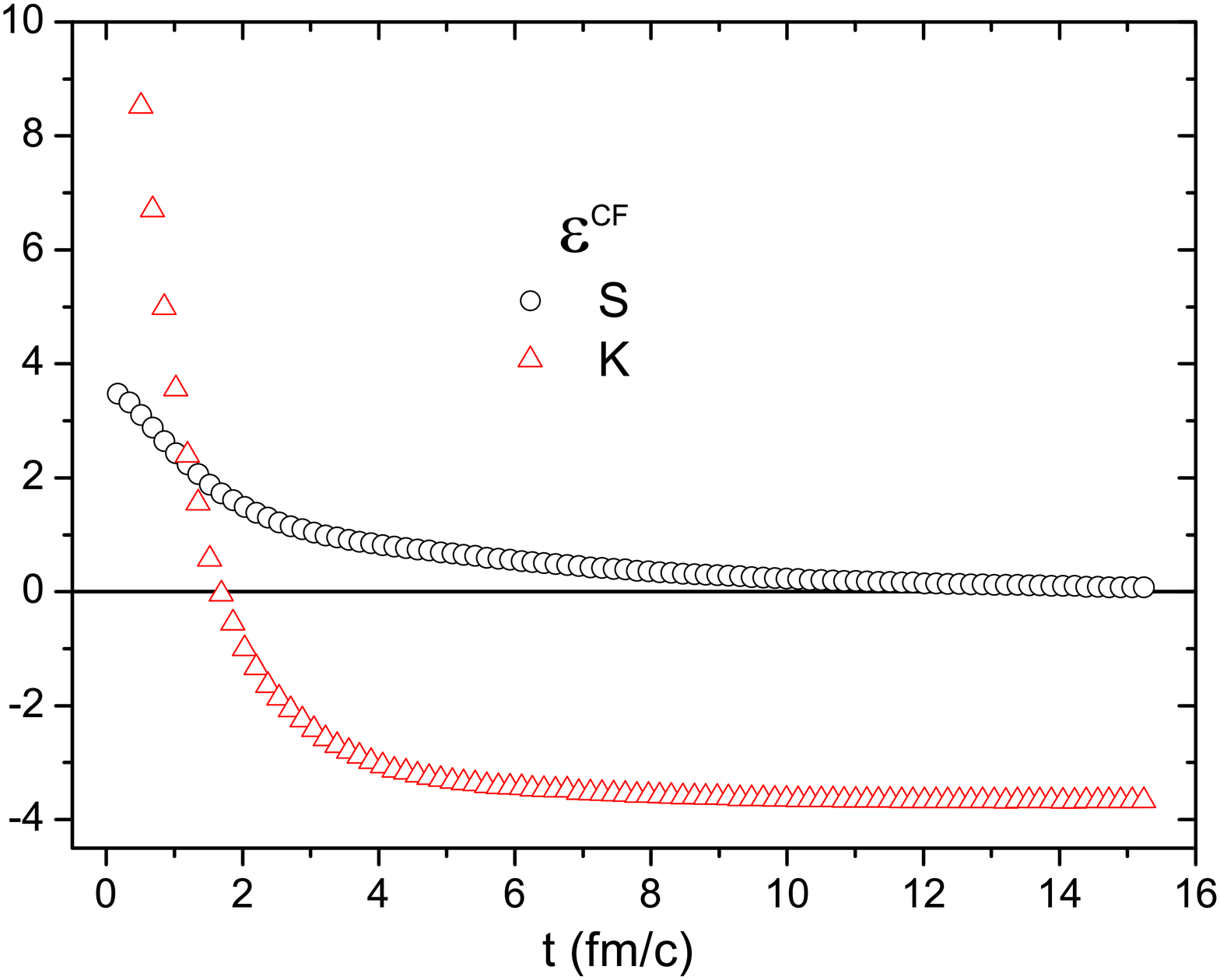}}
\caption{ (color online)
The time development of the kurtosis, $K$, and skewness, $S$,
of the distribution of the specific energy density in the CF,
$\varepsilon^{CF}$, for central Pb+Pb collisions
at $\sqrt{s_{NN}}=2.76$ TeV. The skewness is positive,
decreasing towards zero with time, indicating a longer high energy tail,
especially at early stages where we have more cells with high energy
QGP, and high kinetic energy as discussed in ref. \cite{cser}.  The kurtosis is
initially positive indicating a high energy density QGP with small
spread but decreases with time  and it is becoming negative at 2 fm/c.
This is caused by populating low energy density supercooled states
at later stages.
}
\label{fig7}
\end{center}
\end{figure}

Usually this happens only in very few cells before the estimated
average Freeze Out (FO) time (less than 10\% of the cells), but we did
continue the CFD calculations well into the supercooled state when the
zero-pressure cells amounted to 30-40\% of the total volume.

In some of the thermodynamical variables this change is exhibited
by a change of the development trend line. At the same time the
energetic characteristics of the EoS are realistic, so the overall
development and the basic quantities are well estimated even during the
supercooled stage.

\section{Skewness and Kurtosis}
In the present work we assume a single QGP phase. We study the features
the energy density and net baryon density variations exclusively arising
from the fluid dynamics. This dynamical change is not observable directly,
only at the freeze out hypersurface which is simultaneous or close to the
hadronization. At this point the hadronization may significantly modify the
statistical fluctuations \cite{laszlo1204}. This work evaluates the fluctuations from the
previous fluid dynamical evolution.
In this part we study the skewness and kurtosis of the specific energy
density and the baryon density according to Eq.~(\ref{skewness})
and Eq.~(\ref{kurtosis}).
In Fig. \ref{fig6}, we can see that the kurtosis
of the baryon density
calculated in different frames are similar,
the kurtosis is negative at first and then turns to be positive around
$t=13$ fm/c.
The skewness is allover positive.
 At $t=4$ fm/c, both densities have a minimum, but the skewness value
in the local rest frame is almost twice larger than in the calculational frame.
This indicates that the contribution of the flow
makes the distribution close to Gaussian.
\begin{figure}[ht]  
\begin{center}
\resizebox{0.9\columnwidth}{!}
{\includegraphics{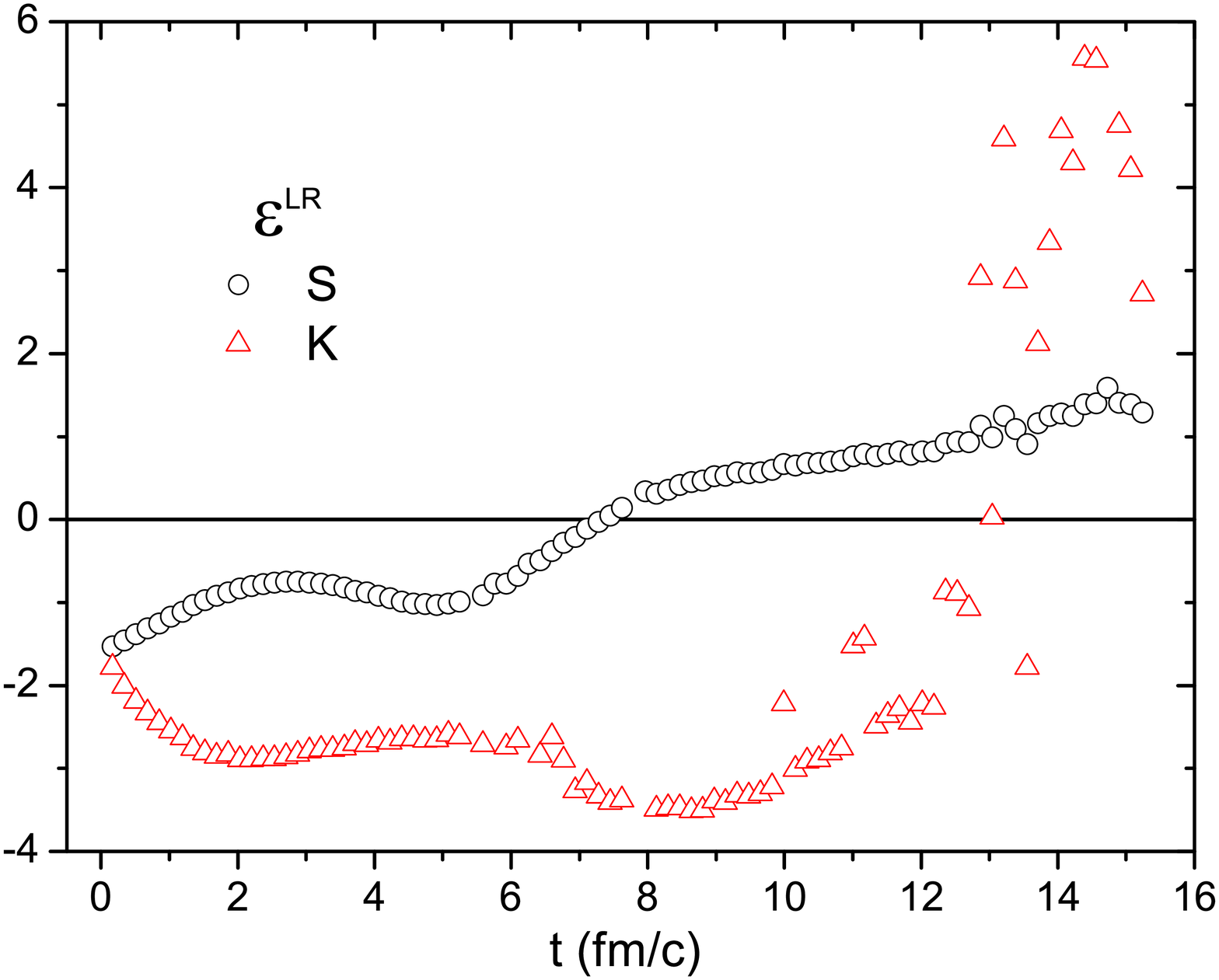}}
\caption{ (color online)
The time development of the kurtosis, $K$, and skewness, $S$,
as in Fig. \ref{fig7}, but for the invariant scalar
specific energy density, $\varepsilon^{LR}$, distribution.
In case of the invariant scalar specific energy distribution,
the skewness changes sign at 7 fm/c.
Unlike in Fig.  \ref{fig7} the skewness is initially
negative indicating a wider spread in the invariant energy density
distribution, which can also be attributed to the presence
of QGP.
}
\label{fig8}
\end{center}
\end{figure}

In Fig. \ref{fig7} we can see that the kurtosis
of the specific energy in CF,
 $\varepsilon^{CF}$,  changes
sign from positive to negative, while the skewness is always
greater than zero.
This is not the same when compared to the results calculated in the local
rest frame, which are shown
in Fig. \ref{fig8}, where skewness changes sign
from negative to positive.

The change of statistics with the changes of the EoS is best seen in
Fig.~\ref{fig8} for the invariant specific
energy distribution  $\varepsilon^{LR}$, although the amplitude of the
change is small. The quantity $\varepsilon^{LR}$, does not include the
contribution from the kinetic energy, thus it is the best measure
of changes in the EoS.  The changes are still observable
contrary to
the fact that the QGP to HM transition is not included in the EoS,
but we still have a transition from ideal QGP to the supercooled,
zero pressure QGP, where the energy density decreases with the
decreasing bag constant.
\begin{figure}[ht]  
\begin{center}
\resizebox{0.9\columnwidth}{!}
{\includegraphics{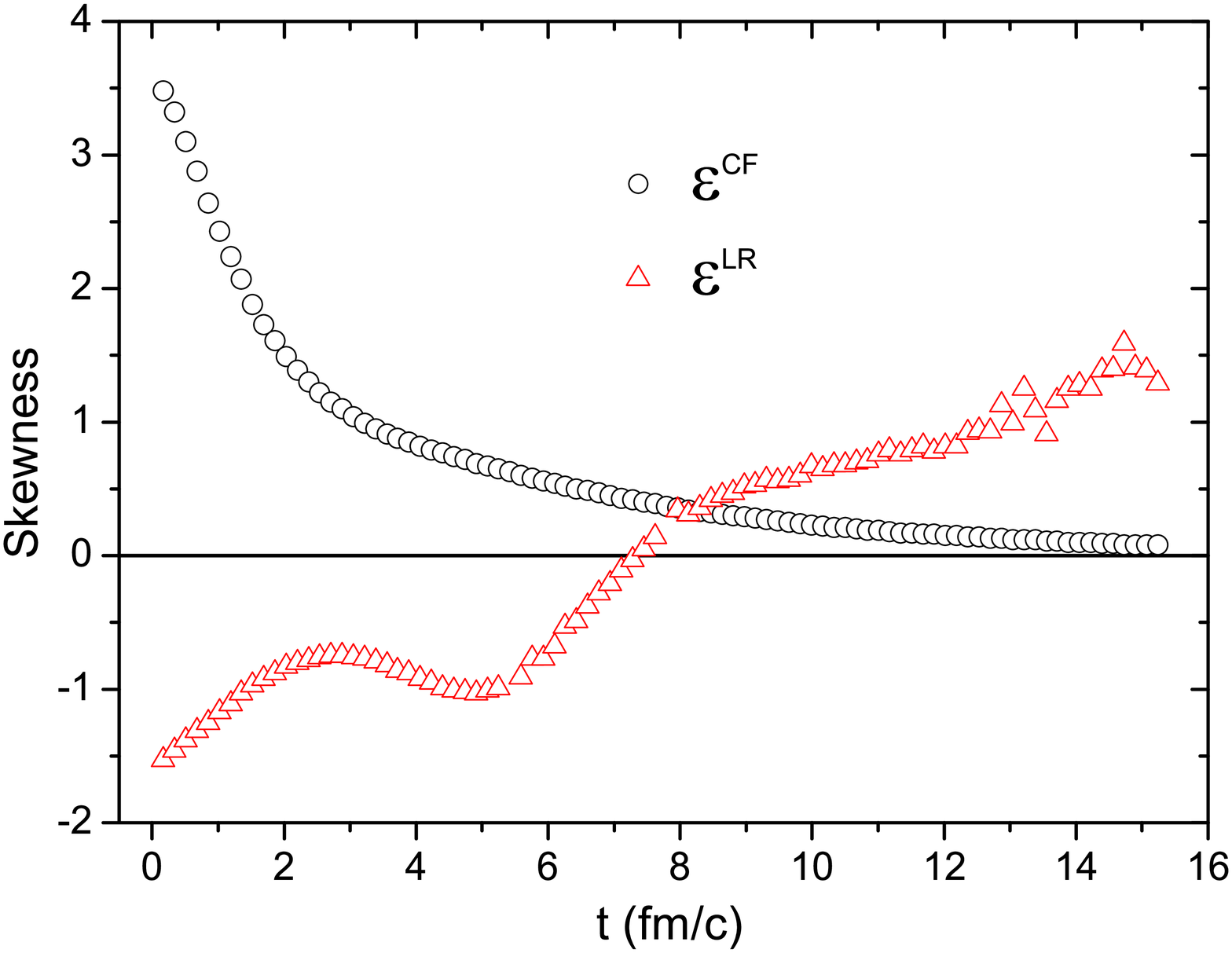}}
\caption{ (color online)
The time development of the skewness, $K$, for the CF specific
energy density, $\varepsilon^{LR}$, and the invariant scalar
specific energy density, $\varepsilon^{LR}$, distribution.
The invariant scalar specific energy distribution changes sign
at the typical FO moment, 8 fm/c.
}
\label{fig9}
\end{center}
\end{figure}

It is interesting to compare the skewness of the two types of energy
densities in Fig.~\ref{fig9}, $\varepsilon^{CF}$ and $\varepsilon^{LR}$.
While $\varepsilon^{CF}$, which contains all kinetic energy, has always a
positive skewness, i.e. the distribution is more uniform.  the
internal energy, $\varepsilon^{LR}$, which increases gradually, has
initially negative skewness, meaning a more spread out, fluctuating
distribution, which becomes positive later, actually at and after
the estimated FO time, when the matter is supercooled and characterized
by zero pressure and reduced Bag constant. We can also note that the cross
point of the skewness in CF and in LR is around 8 fm/c, which is the same
as in Fig.~\ref{fig2}, where the trajectories of the Local Rest energy
and the Kinetic energy cross.
The experimental observations do not show clearly the expected phase transition
behavior \cite{laszlo1204}. That is why we study alternative origins of the fluctuations. The
skewness and kurtosis changes are of the same order of magnitude as from critical
fluctuations, therefore both effects should be studied.

In a multi-module or hybrid-model construction
(e.g where the PIC hydro stage is matched  to a parton and hadron
cascade model PACIAE), the flow features are matched \cite{Yun10}
to a subsequent dynamical model which describes dynamical, non-equilibrium,
rapid hadronization. These types of models can describe realistically
the statistical properties and a dynamical phase transition, which
provide the hadron distribution in the final stage.
This stage would then explicitly describe the random fluctuations
arising from the phase transition also.

In ref. \cite{Daimei-CPOD} a mixed particle method is
introduced, which could separate the fluctuations arising from local
critical fluctuations.
The mixed events are actually eliminating two particle
correlations, and only the single particle distributions
remain. Thus for central events these are mainly local
correlations which may arise from local fluctuations caused
by energy and baryon charge clustering in a phase transition.
The method separates the consequences of such correlations.

This method can be used both in hybrid model calculations and
in experiments, to separate the fluctuation effects from the
collective flow and the phase transition dynamics.

\section{Conclusions}

We studied in a fluid dynamical model the time development of spatial
distribution averages and variances of thermodynamical variables
in high energy heavy ion collisions. We assumed an Equation of State for
Quark Gluon Plasma only, but including supercooled QGP also where the
pressure is dropped to zero. We studied central collisions, without initial
state fluctuations to minimize fluctuations arising from complex
anisotropic flow patterns. Including the supercooled QGP we could
describe the late stages of the collisions expanding up to the phase
transition boundary \cite{Cleymans} and beyond (Fig. \ref{fig1}).
In addition to the average temperature and chemical potential we also
studied the developments of average energy densities and baryon densities
as well as their variances (Figs.
\ref{fig2},\ref{fig3},\ref{fig4},\ref{fig5}).
The dynamical developments of these variables showed the expected,
monotonic dynamical behavior, even beyond the physical FO times,
where we overstretched the applicability of the fluid dynamical model.

Our FD model did not include the rapid hadronization and the random
generation of hadrons, which would generate critical fluctuations
in the vicinity of the critical point of the phase transition.

Interestingly the higher statistical moments, the
Kurtosis and Skewness still showed a non mono\-tonic behavior (Figs.
\ref{fig6}, \ref{fig7}, \ref{fig8}, \ref{fig9}).
We presented these for the highest energy collisions, where
the applicability of the applied EoS is the least questionable.
These higher moments of specific energy densities show changing
signs of the Kurtosis and Skewness.

Apart of the fact that the FD model provides a spatial distribution,
Fig. \ref{fig8} shows that the variation of higher moments is also sensitive
to the FO time. This time cannot be securely determined
from within the FD model, and the final hadronization and freeze-out
should be described by the last model stage of the hybrid model.
Thus, we did not study the excitation function of the higher moments,
because an arbitrary choice of the FO time may result in different
results. On the other hand the time dependence of higher moments obtained
here (Figs. 6-9) is different from the straightforward expectation
arising from a dynamical phase transition \cite{laszlo1204}.

The spatial fluctuations of specific energy and baryon charge density
certainly influence the final baryon charge multiplicity and specific
energy distributions. In the absence of other dynamical effects
during hadronization the fluctuations of the fluid dynamical densities
will be inherited by the corresponding final measurable quantities.

The present calculations show that in dynamical systems, even with
the least initial variation, strongly varying higher statistical
moments may develop. Thus the effects of the FD expansion
and of the final hadronization and freeze out should be separated.
This can be done in theoretical hybrid models by evaluating separately
both effects, and in experiments by mixed event methods
(e.g. ref. \cite{Daimei-CPOD}) or more specific correlation measurements.

\end{document}